\documentclass{article}
\usepackage{emulateapj}
\usepackage{amssym}
\usepackage{psfig}
\usepackage{times}

\newcommand{\gta}{\gtrsim}
\newcommand{\lta}{\lesssim}

\newcommand{\avg}[1]{\ensuremath{\langle#1\rangle}}
\newcommand{\ELP}{\ensuremath{E_{\mbox{\scriptsize\sc lp}}}}

\newcommand{\NLP}{\ensuremath{N_{\mbox{\scriptsize\sc lp}}}}
\newcommand{\Nm}{\ensuremath{N_{\mbox{\scriptsize m}}}}
\newcommand{\Tbb}{\mbox{$T_{\mbox{\scriptsize\sc bb}}$}}
\newcommand{\Tc}{\mbox{$T_{\mbox{\scriptsize\sc c}}$}}

\newcommand{\Tmax}{\mbox{$T_{\mbox{\scriptsize max}}$}}
\newcommand{\emax}{\ensuremath{\epsilon_{\rm max}}}
\newcommand{\eV}{\mbox{e\kern-0.35mm{V}}}
\newcommand{\keV}{\mbox{ke\kern-0.35mm{V}}}
\newcommand{\lc}{\mbox{$l_{\rm c}$}}
\newcommand{\sigmaT}{\mbox{$\sigma_{\mbox{\scriptsize\sc t}}$}}
\newcommand{\tauT}{\mbox{$\tau_{\mbox{\scriptsize\sc t}}$}}
\newcommand{\taue}{\mbox{$\tau_{\rm e}$}}
\newcommand{\taup}{\mbox{$\tau_{\rm p}$}}

\makeatletter
\@ifundefined{chapter}{\def\thebibliography#1{
\section*{References}
  \list
  {\relax}{\setlength{\labelsep}{0em}
        \setlength{\itemindent}{-\bibhang}
        \setlength{\itemsep}{\parskip}
        \setlength{\parsep}{0pt}
        \setlength{\leftmargin}{\bibhang}}
    \def\newblock{\hskip .11em plus .33em minus .07em}
    \sloppy\clubpenalty4000\widowpenalty4000
    \sfcode`\.=1000\relax}}%
{\def\thebibliography#1{
  \list
  {\relax}{\setlength{\labelsep}{0em}
        \setlength{\itemindent}{-\bibhang}
        \setlength{\itemsep}{\parskip}
        \setlength{\parsep}{0pt}
        \setlength{\leftmargin}{\bibhang}}
    \def\newblock{\hskip .11em plus .33em minus .07em}
    \sloppy\clubpenalty4000\widowpenalty4000
    \sfcode`\.=1000\relax}}

\newlength{\bibhang}
\let\@internalcite\cite
\def\cite{\@ifstar{\citey}{\citefull}}
\def\citefull{\def\astroncite##1##2{##1\ ##2}\@internalcite}
\def\citey{\def\astroncite##1##2{##1\ (##2)}\@internalcite}
\def\citeyear{\def\astroncite##1##2{##2}\@internalcite}
\def\citename{\def\astroncite##1##2{##1}\@internalcite}
\def\@citex[#1]#2{\if@filesw\immediate\write\@auxout{\string\citation{#2}}\fi
  \def\@citea{}\@cite{\@for\@citeb:=#2\do
    {\@citea\def\@citea{; }\@ifundefined
       {b@\@citeb}{{\bf ??}\@warning
       {Citation `\@citeb' on page \thepage \space undefined}}%
{\csname b@\@citeb\endcsname}}}{#1}}

\def\@cite#1#2{#1\if@tempswa #2\fi} 
\def\@biblabel#1{}

\def\astroncite#1#2{#1\ #2}
\makeatother

\begin{document}

\title{Self-Consistent Thermal 
Accretion Disk Corona Models for Compact Objects: I. Properties of the
Corona and the Spectrum of Escaping Radiation}
\author{James B. Dove\altaffilmark{1,2}, J\"orn Wilms\altaffilmark{3,1}, and 
Mitchell C. Begelman\altaffilmark{1,2}}

\altaffiltext{1}{JILA, University of Colorado and National Institute of
Standards and Technology, Campus Box 440, Boulder, CO 80309-0440\\
\{dove,mitch\}@rocinante.colorado.edu.}

\altaffiltext{2}{Department of Astrophysical, Planetary, 
and Atmospheric Sciences, \\ University of
Colorado, Boulder, Boulder, CO 80309-0391.}

\altaffiltext{3}{Institut f\"ur Astronomie und Astrophysik,
Abt.~Astronomie, Waldh\"auser Str. 64, D-72076 T\"ubingen, Germany\\
wilms@astro.uni-tuebingen.de}

\begin{abstract}
We present the properties of accretion disk corona (ADC) models, where the
radiation field, the temperature, and the total opacity of the corona are
determined self-consistently.  We use a non-linear Monte Carlo code to
perform the calculations. As an example, we discuss models where the corona
is situated above and below a cold accretion disk with a plane-parallel
(slab) geometry, similar to the model of Haardt and Maraschi.  By
Comptonizing the soft radiation emitted by the accretion disk, the corona
is responsible for producing the high-energy component of the escaping
radiation.  Our models include the reprocessing of radiation in the
accretion disk.  Here, the photons either are Compton reflected or
photo-absorbed, giving rise to fluorescent line emission and thermal
emission.  The self-consistent coronal temperature is determined by
balancing heating (due to viscous energy dissipation) with Compton cooling,
determined using the fully relativistic, angle-dependent cross-sections.
The total opacity is found by balancing pair productions with
annihilations.  We find that, for a disk temperature $k\Tbb \lta
200$\,\eV, these coronae are unable to have a self-consistent temperature
higher than $\sim 120$\,\keV\ if the total optical depth is $\gta 0.2$,
regardless of the compactness parameter of the corona and the seed opacity.
This limitation corresponds to the angle-averaged spectrum of escaping
radiation having a photon index $\gta 1.8$ within the $5$\,\keV\ -
$30$\,\keV\ band.  Finally, all models that have reprocessing features also
predict a large thermal excess at lower energies.  These constraints make
explaining the X-ray spectra of persistent black hole candidates with ADC
models very problematic.
\end{abstract}

\section{Introduction}\label{introsec}
The high energy spectra from Galactic black hole candidates (BHCs) and
radio quiet active galactic nuclei (AGN) are similar. Both can be described
roughly by a power law, with a photon index of $1.5 \sim 2.0$ that extends
up to $\sim 100$\,\keV, modified by an exponential cutoff
(\cite{grebenev93,gilfanov93,maisack93,wilms97a}, and references therein).
The spectra of many compact objects also show features due to reprocessing
of X-rays with moderately cold matter, including a fluorescence line due to
iron and a Compton reflection hump
(\cite{guilbert88,pounds90,nandra91,fabian94}, and references
therein).

Comptonization of radiation by a semi-relativistic plasma, in the form of a
corona associated with an accretion disk, may be the principal mechanism
that produces the high-energy portion of these spectra. However, even
within the context of these models, the geometry and physical properties of
the corona are still being debated.  One popular model involves an
optically thick, cold ($kT \lta 1$\,\keV) accretion disk and a hot
accretion disk corona (ADC), often having a plane-parallel (slab)
configuration (Galeev, Rosner, \& Vaiana, 1979;
\cite{sunyaev80,haardt91}, hereafter HM91; \cite{haardt93a},
hereafter HM93; Haardt, Maraschi, \& Ghisellini, 1994, 1997;
\cite{nakamura93,hua95,stern95a,poutanen96a}, and references therein).  One
nice feature of the ADC models is that the accretion disk serves a double
role in being the source of the seed-photons that are Comptonized by the
corona and being the medium responsible for the reprocessing and reflection
features.  It is commonly assumed that a substantial fraction of the
gravitational accretion energy is dissipated directly into the corona,
although the physical processes by which the corona is heated are still
unknown.

Recent theoretical work provides a heuristic framework for understanding
why accretion disk coronae should form.  \citename{balbus91a}
(\citeyear{balbus91a}; see also \cite{hawley94a}, and references therein)
have found a powerful MHD instability which, in its nonlinear development,
tends to amplify the turbulent magnetic pressure to a level on the order of
the gas pressure.  Buoyancy of the magnetic flux tubes will tend to make
them rise out of the disk interior (\cite{galeev79,tout92a}), into a region
where they dominate the ambient gas pressure.  As is believed to occur
in the solar corona, this type of magnetic field evolution gives rise to a
field structure in which large-scale magnetic reconnection is inevitable,
releasing energy which can heat the corona
(\cite{hawley96a,balbus96a}).  More recent work shows that the magnetic
field energy density is increased in low density regions (i.e., the outer
disk atmosphere), possibly allowing for direct coronal heating
(\cite{stone96a}).

Alternate accretion disk models also have been able to explain the
high-energy spectra of X-ray binaries. One common alternative model
involves an optically thin, hot accretion disk
(\cite{shapiro76,kusunose95,white90,luo94,narayan95b}, and references
  therein; \cite{Chen95}, and references 
therein).  However, these models, unlike the ADC models, must include an
additional reprocessing medium (e.g. a stellar wind or an outer accretion
disk) in order to explain the observed reflection features.

Due to the coupling between the radiation field and the coronal plasma,
solving the non-linear radiative transfer problem for accretion disk
coronae is very complicated.  Through Comptonization, the properties of the
radiation field depend on the corona's temperature and opacity.  Since
Compton cooling is the dominant cooling process, the temperature of the
corona, in turn, depends on the radiation field.  Additionally, the corona
and radiation field are coupled to each other through the processes of pair
production and pair annihilation.  Reprocessing of radiation within the
accretion disk, where the radiation is either Compton-reflected back into
the corona or photo-absorbed, also complicates the problem.

For accretion disk corona models, the high-energy spectrum of escaping
radiation depends primarily on the geometry of the system, the Thomson optical
depth of the corona, $\tauT$, and the temperature of the corona,
$\Tc$.  For a given geometry, the last two parameters typically are
allowed to vary independently until a good fit to the data is achieved.
However, an ADC system with a given \Tc\ and \tauT\ is not necessarily a
self-consistent model.  For many geometries, reprocessing of radiation from
the corona within the cold accretion disk allows the accretion disk to have
a large flux of thermal radiation even if most of the gravitational energy
is dissipated within the corona (\cite{haardt91}).  As we will show 
for the specific case of slab geometries, a high flux of soft photons
results in high Compton cooling rates, and consequently models with
modest optical depths ($\tauT \gta 0.2$) can only have temperatures $\Tc
\lta 120$\,\keV.  Therefore, previous ADC models that give acceptable fits to
the data of BHCs (\cite{haardt93a,titarchuk94b}) are not 
physically realizable.  For example, the spectrum of Cyg X-1 has recently
been described by an ADC model, with a slab-geometry, having a temperature
of $\Tc \sim 150$\,\keV\ and an optical depth of $\tauT \sim 0.3$
(\cite{haardt93a}).  As we discuss below, this is not a self-consistent
combination.

\citename{haardt91} (\citeyear{haardt91}; \citeyear{haardt93c};
\cite{haardt93b}) 
developed the first thermal Comptonization model in which the coupling
between the radiation field and the coronal plasma is taken into
account. Reprocessing of radiation within the cold accretion disk also is
included in their model, and a self-consistent temperature of the corona is
determined by equating Compton cooling with viscous heating.  As we will
discuss in \S\ref{prevmodsec}, this model uses a hybrid of Monte Carlo (MC)
and analytical techniques.  However, several of the approximations used in
calculating the radiation field limit the accuracy and validity of these
calculations.

In this paper we present our ADC model that uses a non-linear Monte Carlo
(NLMC) code adapted from Stern's ``Large Particle'' Monte Carlo technique
(\cite{stern85,stern95a}, and references therein).  Although we only
consider the slab geometry in this paper, more complicated geometries can
be modeled and will be the focus of future papers. Rather than propagating
photons through a ``background" medium one at a time, as is the case with
linear Monte Carlo methods (Pozdnyakov, Sobol', \& Sunyaev, 1983), all
types of particles are propagated in parallel (for this paper, the types of
particles included are photons, electrons, and positrons).  In addition,
the energetics of the system, including the Compton cooling rate, the
pair-production rate, and the electron/positron annihilation rate, are
calculated numerically. Thus, the pair opacity, the temperature of the
corona, and the radiation field can be computed self-consistently. These
quantities are computed {\em \it locally}, giving rise to models with
non-uniform temperature and/or density distributions (see discussion in
\S\ref{resultssec}).  Fully relativistic and angle-dependent cross sections
are used, and there is no need to make any radiative transfer
approximations.  The pair production rates and Compton cooling rates are
computed using the self-consistent, angle-dependent radiation field rather
than analytic approximations.

The remainder of the paper is organized as follows.  In \S\ref{prevmodsec},
we discuss the limitations of the computational methods used in previous
ADC models. In \S\ref{nlmcsec}, we summarize the ``large particle'' Monte
Carlo method and describe our modifications that make it more applicable to
thermal accretion disk corona models.  In \S\ref{modelsec}, we describe our
ADC model and note the differences between it and previous models.  In
\S\ref{resultssec}, we present the results of our simulations, and in
\S\ref{concsec}, we summarize our results and give our plans for future
work.

\section{Previous ADC Models and Computational Techniques}\label{prevmodsec}
To date, most thermal Comptonization models use either the analytical
methods pioneered by \cite*{lightman80} and \cite*{sunyaev80} (and expanded
upon by \cite{sunyaev85} and \cite{hua95}), numerical methods of solving
the kinetic equations
(\cite{zdziarski85,svensson87,lightman87,coppi92,poutanen96a}, and
references therein), or ``linear'' Monte Carlo (MC) methods
(\cite{pozdnyakov83}; \cite{gorecki84}). Since the properties of the corona
have to be assumed before the radiation field is computed with an
analytical method, it is uncertain whether the properties of the corona are
self-consistent with the radiation field. In addition, analytical solutions
can be found for only a small number of simple coronal geometries (most
importantly spherical and plane parallel geometry) and it does not appear
possible to perform the computations using the fully relativistic,
angle-dependent cross sections (\cite{hua95}). Also, no analytic method has
been presented that models the reprocessing of radiation in the accretion
disk while taking into account the anisotropy of the radiation field within
the corona. Thus, if reprocessing is important in the source, ``reflection
features'' have to be computed using a different computational method and
added to the escaping radiation field. With this approach, subsequent
interactions between the reflected component of the radiation field and the
ADC cannot properly be taken into account.

Linear MC methods, where one photon at a time is propagated through a fixed
medium, have to specify the properties of the corona before simulating the
radiation field; as it is the case with the analytical models, the
specified properties of the corona often are not self-consistent.  In
addition, since linear MC methods follow one photon at a time, simulating
photon-photon pair productions and annihilations must be done analytically,
and pair opacity must be  approximated by interfacing these
semi-analytic calculations with the MC computations (HM93).  

As discussed in the introduction, \citename{haardt91} (HM91, HM93)
developed an improved thermal Comptonization model for plane-parallel
accretion disk coronae by taking the coupling between the accretion disk
and the corona into account. The temperature of the corona is calculated by
balancing the total Compton cooling rate with the assumed heating rate
(which is determined by the coronal compactness parameter, a free
parameter).  All of the gravitational accretion energy is dissipated
uniformly into the corona.  Therefore, all the energy emitted by the cold
accretion disk is supplied by irradiation from the corona.  The total
opacity is calculated by balancing photon-photon pair productions with
annihilations.

The models by HM91 and HM93 are the first models in which the coronal
temperature and opacity are not independent free parameters.  However, due
to the analytical approximations, these models have several shortcomings:
(1) Since HM93 use the Thomson cross section in the electron's rest frame
rather than the correct Klein-Nishina cross section, a Wien cutoff, rather
than a self-consistent spectral shape, is used to approximate the
spectrum for energies $\epsilon \gta k\Tc$, where \Tc\ is the
temperature of the corona.  This approximation leads to an underestimation
of the pair-production rates since the high-energy tail of the
self-consistent spectrum is harder than a Wien cutoff (\cite{stern95b}).
(2) Additional uncertainties arise because the Compton cooling rate is
calculated by approximating the spectrum of radiation as a power law with a
Wien cutoff. (3) HM93 neglect multiple reflections, line features, and
subsequent interactions with the corona by reflected photons.  Very
recently, \cite*{haardt97a} 
have modified the HM93 model such that the high-energy roll-over is
calculated more accurately by using a fully relativistic kernel for
isotropic unpolarized radiation (see \cite{haardt94} for a detailed
discussion).

Using an iterative scattering method, \cite*{poutanen96a} solve the
radiative transfer problem for ADC models, in which Compton scattering,
photon-photon pair production, pair annihilation, bremsstrahlung, and
double Compton scattering are taken into account.  In addition, the
relativistic cross sections are used, the escaping spectrum of radiation
can be determined for any direction, and the reprocessing of coronal radiation
in the accretion disk is taken into account using a reflection matrix.
The main limitation of this method is that it is one dimensional, and
therefore complicated geometries cannot be considered. Another
limitation of this method is that solutions can be found only for
small optical depths ($\tauT \lta 1$, depending on 
the coronal temperature; see \cite{poutanen96a} for more details).

\section{The Non-Linear Monte Carlo Code}\label{nlmcsec}
\subsection{Overview}
To study the non-linear problem of radiative transfer in an ADC, we have
modified the ``Large Particle'' Monte Carlo (LPMC) method, developed by
\citename{stern85} (\citeyear{stern85,stern88}) and \citename{stern95a}
(\citeyear{stern95a,stern95b}).  Unlike most Monte Carlo 
codes used for accretion disk coronae, where photons are propagated through
a ``background'' medium one at a time, the LPMC code simulates all
particles (photons, electrons, positrons) in parallel (i.e.,\ 
simultaneously).  The advantage of propagating all particles in parallel is
that there is no ``background'' medium; photons can interact with
electrons and like particles can
interact with each other.  Because the coupling between the radiation field
and the plasma is taken into account by directly simulating all particles
in parallel, the non-linear radiation-transfer problem is solved
self-consistently.

One challenge of using Monte-Carlo numerical techniques for simulating
X-ray spectra is to obtain good statistics with models requiring only a
modest amount of CPU time.  Since we are simulating high-resolution spectra
that span many decades in energy, it is not possible to simulate the
particles by just interpreting any particle within the code as a particle
in the real physical system. If this approach were used, a great deal of
CPU time would be wasted by tracking the low-energy portion of the photon
spectrum.  To deal with this problem, \cite*{stern85} developed a numerical
technique using ``large particles'' (LPs).  Each LP represents a group of
identical particles having the same particle type (photon, electron, or
positron), energy, velocity, and position.  All LPs have the same total
energy $\ELP=\epsilon \omega$, where $\omega$ is the statistical weight of
the LP (proportional to the number of physical particles represented by the
LP) and $\epsilon$ is the energy of each of the physical particles within
the LP.  With all LPs having the same total energy $\ELP$, the largest
number of LPs and thus the best statistics will be in the spectral region
where the spectral energy density is highest.

The method of using LPs can be considered to be a generalization of the
``particle splitting'' techniques which are routinely used to improve
statistics in linear Monte Carlo simulations (\cite{pozdnyakov83}).  
However, an escape probability formalism is not used.
Instead, the LPs are split due to interactions within the system.  When an
LP undergoes interactions, all of the physical particles represented by the
LP are required to experience the identical consequences. Since the total
energy of an LP is fixed, $\epsilon \propto 1/\omega$, the statistical
weight must change as the energy of the physical particles, $\epsilon$,
changes due to the interaction. This means that the number of physical
particles represented by the LP will change. A consequence of this is that
each interaction requires the elimination or creation of new LPs such that
energy, momentum, and particle number are conserved in a statistical
sense. For example, if $\epsilon$ doubled due to a collision, then the
statistical weight of the LP would be reduced by a factor of two. 
In order to conserve particle number, an additional LP, having the same
properties (energy, weight, direction, etc.) as the first LP, would be
created.  The reader is referred to \cite*{stern95a} for more details.

The most important feature of the NLMC method is that all of the LPs are
propagated in parallel.  To do this, each LP is propagated during each
time-step, which is chosen to be small enough such that $\lta 5$\% of the
LPs undergo an interaction.  While each LP propagates, the properties of
the system (e.g., optical depth and temperature) used for
determining the collision probabilities are determined from the state of the
system at the end of the previous time-step.  All of the properties are
updated after the last LP is propagated.  By using such small time-steps,
it is ensured that the properties of the system will not change
significantly during the time-step, and consequently each LP will ``see''
essentially the same configuration.  Small time-steps also allow the
properties of the system to evolve smoothly.

For this paper, the type of particles that are simulated are photons,
electrons and positrons (although our numerical routine also is capable of 
simulating other particles).  Although we do not simulate protons because
the cross section for Compton scattering is much smaller than that of the
electrons and positrons, we specify their initial distribution and enforce
charge neutrality such that the initial distribution of electrons is equal
to the specified proton distribution.  During the simulation, we require
that $n_{\rm e} - n_+ = n_{\rm p}$, where $n_{\rm e}$, $n_{+}$, and $n_{\rm
p}$ are the number density of electrons, positrons, and protons,
respectively.

The physical interactions that are taken into account are: Compton
scattering, photon-photon pair production, and pair annihilation.  The
numerical code is also capable of simulating synchrotron radiation and
$e^-\,e^-$- and $e^-\,e^+$-Coulomb interactions; for simplicity, however,
we do not consider these processes in this paper.  For the reprocessing of
radiation in the accretion disk, Compton scattering and photo-absorption,
resulting in fluorescent line emission and thermal emission, are
considered.  Simulations of all ``two-body'' processes are carried out using
the fully relativistic, angle-dependent, QED cross sections as well as 3-D
relativistic kinematics (\cite{akhiezer65}).  Continuous processes are
simulated through discrete ``collisions'' using the same techniques as for
the `two-body' processes. However, the cross sections that determine the
collision rates are chosen such that the time-average rate of energy losses
agrees with the continuous, analytical cooling rates.

In order to allow for non-uniform corona models, the corona is divided into
spatial cells. For plane-parallel geometry these cells are uniform layers
with a thickness $\Delta z$ (although it is possible for $\Delta z$ to
depend on the cell number). Since the radiation field and the properties of
the corona within each cell are assumed uniform, $\Delta z$ is chosen such
that each cell is optically thin.  However, as discussed in
\S\ref{statsec}, if the statistical weight within a cell is too small,
statistical fluctuations become problematic.  We find that ten spatial
cells allows for good resolution of the vertical dependence of the
temperature, opacity, and radiation field while also allowing for
acceptable statistics within each cell.

\subsection{Modifications to the NLMC Code}
The LPMC Code was originally invented to model pair cascades in
relativistic pair plasmas (\cite{stern85}).  We made several modifications
to the code to make it more suitable for studying accretion disk coronae.
These modifications include the use of thermal pools and the treatment of
radiation reprocessing in the accretion disk.

\subsubsection{Thermal Pools}
For thermal accretion disk corona models, it is not necessary to treat all
of the electrons and positrons as LPs since it is assumed that most
of them are thermalized.  Instead, for each space cell, a thermal
electron ``pool'' and a positron ``pool'' are used.  The pools are
characterized by having a statistical weight, $w(i)$, and a temperature,
$\Tc(i)$, where $i$ is the cell number in the corona.  Since bulk
motion is not considered in this paper, the velocities of the individual
``particles'' within the pool are statistically determined from a
relativistic Maxwellian distribution appropriate for the pool's temperature.
Within each cell, we assume that the density distribution of the pool
`particles' is uniform.  The temperature of the positron pool
is assumed to be equal to the temperature of the electron pool.  At the
beginning of a simulation, all electrons and positrons are assumed to be
thermalized and are put into their respective pools.  When an LP is
interacting with the pool (e.g., by Compton scattering with a photon LP), a
`mini-particle,' with a weight equal to the weight of the interacting LP,
is drawn from a relativistic Maxwellian distribution.  If, as a result of
the interaction, the `mini-particle' ends up with an energy greater than a
cut-off energy $\emax(i)$, then the `mini-particle' is ejected
from the pool and converted to an LP.  For this paper, we set
$\emax(i)=8k\Tc(i)$ such that over 99\% of the electrons
and positrons are thermalized, and only the very energetic particles are
treated as LPs.  Alternately, after an electron or positron LP
interaction or following a pair-production event, the electron or positron
LP is inserted into the corresponding pool if its energy is less than the
cut-off energy. Once an LP is inserted into the pool, its identity is lost,
but the pool gains a statistical weight and an energy equal to that of the
LP.

The use of thermal pools makes the computations much more efficient because
it is not necessary to simulate all low-energy particle interactions within
the pool. These physical low-energy interactions thermalize the electrons
and positrons, and we already have assumed the pools to be thermalized.
The probability that an LP interacts with the pool is determined in the same
fashion as for LP-LP interactions, but the total weight of the pool is used
instead of the individual weight of the `target' LP.  The only pool-pool
interactions we simulate are annihilations between the positron pool and
the electron pool.  Pool annihilations are treated by dividing the positron
pool into 100 `mini-particles,' each having a weight equal to the total
weight of the pool divided by 100.  Each positron mini-particle is given a
velocity, drawn from a relativistic Maxwellian distribution, and a random
direction of motion.  Each mini-particle is then propagated in time.
For each annihilation event, a statistical weight $w_{\rm e}/100$ is taken
out of both the electron pool and the positron pool, while an equivalent
weight is inserted in the form of photon LPs.

\subsubsection{Thermal Structure}
The thermal structure of the corona is given by the distribution of the
pool temperature of each cell, and the non-thermal structure is given by
the distribution of electron and positron LPs.  The temperatures of the
pools adjust in time due to energy gains and losses through LP
interactions, pool annihilations, and external energy sources (which we are
associating with dissipation of accretion energy in this paper).  The
equilibrium pool temperature is given implicitly by requiring
\begin{equation}
H(z) - C({\rm T,n_{\rm e,+},U_{\rm rad}}) = 0,
\end{equation}
where $H(z)$ is the external heating rate per unit volume, $z$ is the
height above the accretion disk, $C$ is the {\it net} cooling rate per unit
volume, $n_{\rm e,+}$ are the number densities of electrons and positrons, and
$U_{\rm rad}$ is the energy density of the radiation field.  Compton
scattering with low-energy photons is the dominant method of corona cooling
and external energy deposition is the dominant source of corona heating.

\subsubsection{Disk Reprocessing}
The reprocessing of radiation within the cold accretion disk is simulated
using a linear Monte Carlo method. Physical processes included are Compton
scattering, photo-absorption, and fluorescence. Since we will apply this
model to the spectra of BHCs that contain an Fe K$\alpha$ line, we we
assume that all elements in the disk are neutral and simulate Compton
scattering off stationary electrons, and the main results described below
are not influenced by this assumption.  In this case, the main changes in
the reflected spectrum are at energies below about 10\,\keV. We use
photoionization cross-sections from \cite*{band:90} and \cite*{verner93}
and solar abundances given by \cite*{grevesse89} (taking into account the
smaller iron-abundance found by \cite{biemont91a} and \cite{holweger91a},
where $[Fe] = 7.54$). Fluorescence in the disk is simulated using the
fluorescence yields given by \cite*{kaastra93}. In order to treat multiple
scattering and absorption/re-emission events, we use a variation of the
method of weights (\cite{pozdnyakov83}).  Photons absorbed during a
scattering event are re-emitted with their energy changed to the energy of
the fluorescence line and their weight reduced by a factor given by the
fluorescence yield.  Energy is conserved by forcing the rate of LPs emitted
as black-body photons and line-photons to equal the rate of absorption
events.  Photons that scatter within the disk more than 50 times are
assumed to be thermalized, leading to the emission of a black-body photon.

\subsection{Statistical Fluctuations}\label{statsec}
Due to statistical fluctuations that arise from using a finite number of
LPs in the simulations (currently we use $\NLP = 65\,536$), the Compton
cooling and pair production rates vary between iterations, which can cause
large temperature and opacity oscillations.  Due to the the strong coupling
between the temperature and the opacity of the corona, intrinsic
fluctuations in the temperature leads to fluctuations in the
opacity (and vica versa).  These oscillations can lead to an escaping photon
spectrum that differs from that of a ``steady-state'' model (even if the
two models have the same ``time-averaged'' properties). In order to reduce
the magnitude of the fluctuations, the temperature of each electron or
positron pool is averaged over the previous four iterations. Averaging over
a larger number of steps leads to over-stable oscillations in both the
temperature and pair-opacity because the corona cannot adjust to changes
in the radiation field quickly enough, causing it to overshoot its
equilibrium values.  An additional improvement is made by averaging the
opacity over the previous ten iterations.  This averaging still allows for
the correct annihilation and pair-production rates, when averaged over all
the iterations, and therefore does not alter the equilibrium pair opacity
values. Occasionally, even when averaging, severe fluctuations will cause
significant temperature changes.  We prevent this occurrence by
artificially preventing the temperature of the pool to change by more than
10\% between iterations.  When the system is in equilibrium, the
fluctuations of the cooling rates are symmetric around the equilibrium
values.  Thus, even though energy is not conserved in the iterations where
large temperature changes are artificially prohibited, the energy of the
system is conserved in a statistical or ``time-averaged'' sense.

An additional source of statistical fluctuations of the pool properties is
due to unlikely LP-pool interactions.  For example, 
there is a very small probability that a Compton-scattered photon
gives all of its energy to a ``pool-electron'' particle, causing the
particle to be ejected from the
pool as discussed above. For optically thin models, where the statistical
weight of the electron pool is small, these ejection events can lead to
large fluctuations in the opacity. In order to reduce the magnitude of
these fluctuations, LPs that interact with a pool are `split' into $\Nm$
mini-LPs, where the mini-LPs have a statistical weight equal to
$1/\Nm$ of the original weight of the LP. Each of the mini-LPs is
then propagated in the usual way, and the final state of the LP and pools
is determined in a statistical sense.  We find that $\Nm = 25$ allows
for adequate statistics without degrading the efficiency of the code too
much.  For models with a total optical depth of the corona $\tau \lta 0.1$,
we let $\Nm=50$.

Using the above methods, the RMS statistical fluctuations of both the
coronal temperature and the total optical depth are $\sigma_{\rm rms} \lta
5$\% after the system has reached its `steady-state' equilibrium values.
We emphasize that, although the simulations are integrated in `time,' the
true time-dependent behavior of the simulations may not be correct since
the statistical methods force the timescales for the changes in the
temperature and opacity to be longer than their proper values.  Thus,
time-dependent simulations are only meaningful on time-scales longer than
the time needed for the corona properties to vary smoothly.  In this paper,
we study only ``steady-state'' models and use ``time'' only insofar as it
measures the approach to an equilibrium.  Thus, the transient behavior
(which arises from simulating one model by starting from the ending point
of a different model) of the radiation field and the coronal properties is
disregarded.  Only after the simulation has reached a steady-state do we
begin recording the spectrum and determining the corona properties.

\section{The Slab Geometry ADC}\label{modelsec}
Although more complicated geometries are easily simulated with our code,
for this paper we consider only the plane-parallel geometry.  In addition,
although we can have non-uniform proton density distributions and
non-uniform heating rates, for this paper we consider uniform models such
that we can compare our results to previous work.  Lastly,
we assume that the protons have the same temperature as that of the
electrons.  In future papers, we will relax all of these simplifications.

In our models, the rest-mass energy density of electrons is several orders
of magnitude lower than the radiation energy density.  Therefore, the
collision rate between photons and electrons is much higher than the
Coulomb collision rate.  The higher Compton collision rate indicates that
the radiation field and the cooling rate of the corona are dominated by
Compton collisions, so we neglect cooling due to thermal bremsstrahlung.
For simplicity, we also neglect synchrotron radiation for this paper.

\begin{figure*}
\centerline{\psfig{file=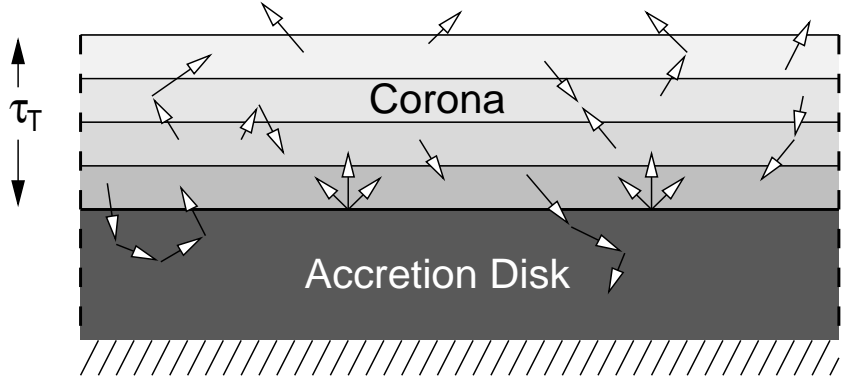,width=0.6\textwidth,angle=-90}}
\caption{Plane parallel (slab) geometry. The accretion disk
is assumed to be optically thick; all photons that are incident
onto the disk are either reflected or photo-absorbed.
The disk is assumed to be symmetric around its mid-plane (indicated by the
hatched region).}\protect\label{fig:slab}
\end{figure*}

In the plane-parallel (slab) geometry, the accretion disk corona is above
and below the optically thick (but geometrically thin) accretion disk, as
shown in Fig.~\ref{fig:slab}.  
For this model, the geometric covering
factor, $\Omega/2\pi = 1$.  Similar to HM93, we assume that all corona and
disk properties are constant with respect to radius and assume azimuthal
symmetry.  The principal parameters of our models are: (1) the seed optical
depth of the corona, \taup, (which does not include the contribution due to
pairs), (2) the local compactness parameter of the corona, \lc, and (3) the
temperature of the accretion disk, \Tbb.  For slab geometry, we define a
{\it local} compactness parameter of the corona as
\begin{equation}
\lc = \frac{\sigmaT}{m_{\rm e} c^3}z_{\rm 0} \Psi_{\rm c},
\end{equation}
where $z_{\rm 0}$ is the scale-height of the corona, $m_e$ is the mass of
the electron, $c$ is the speed of light, and $\Psi_{\rm c}$ is the rate of
energy dissipation per unit area into the corona.  We remind the reader
that $\lc$ is a {\it local} compactness parameter, and is related to
the global compactness parameter used by HM93 by $\lc({\rm local}) =
\lc({\rm global})/\pi$ for cylindrical disk geometry with a uniform
$\Psi_{\rm c}$.  The optical depth of the corona is given by $\tauT =
\int^\infty_{\rm 0}\sigmaT n(z){\rm d}z$, where $\sigmaT$ is
the Thomson cross section and $n(z)$ is the number density of all electrons and
positrons.  

\subsection{Energetics}
We define a local disk compactness parameter as
\begin{equation}
l_{\rm d} = \frac{\sigmaT}{m_{\rm e} c^3}\, z_{\rm 0} F_{\rm d},
\end{equation}
where $F_{\rm d}$ is the flux of radiation emitted by the disk
(erg\,s$^{-1}$\,cm$^{-2}$) in the form of thermal radiation and fluorescent
lines.  Radiation that is reflected off the disk is not included in $F_{\rm
d}$.  The flux of radiation that is absorbed by the accretion disk is given
by
\begin{equation}
F_{\rm abs} = F_{\rm c}^{-}(1 - a),
\end{equation}
where $F_{\rm c}^{-}$ is the flux of downward directed radiation within the
corona just above the accretion disk and $a$ is the albedo of the disk
(averaged over all energies and angles).  We require that the flux of radiation
emitted by the disk equals the sum of the rate of gravitational energy
dissipated within the disk and the flux of downward directed radiation
absorbed by the disk.  This requirement is expressed as
\begin{equation}\label{fbbeq}
F_{\rm d} = (1-f)P_{\rm G} + F_{\rm abs},
\end{equation}
where $P_{\rm G}$ is the rate of gravitational energy dissipated per unit
area, and $f$ is the fraction of this energy that is deposited directly
within the corona.  By defining the total compactness parameter due to
gravitational dissipation to be $l_{\rm G} = (\sigmaT/m_{\rm e}
c^3)\,z_{\rm 0}\,P_{\rm G}$, equation~(\ref{fbbeq}) becomes
\begin{equation}
l_{\rm d} = l_{\rm G}(1-f) + l_{\rm abs},
\end{equation}
where $l_{\rm abs} = (\sigmaT/m_{\rm e} c^3)\,z_{\rm 0}\,F_{\rm
abs}$. Finally, by defining the escaping radiation compactness parameter
as $l_{\rm esc} = (\sigmaT/m_{\rm e}c^3)\,z_{\rm 0}\,F_{\rm
esc}$, the requirement of conservation of energy can be expressed as
\begin{equation}
l_{\rm G} = l_{\rm esc} = \lc + l_{\rm d} - l_{\rm abs}.
\end{equation}
Unlike HM93, we do not assume $f=1$.  Instead, we allow for an intrinsic
disk compactness by setting $(1-f)l_{\rm G} = 1$ for all models.  The
fraction of gravitational energy dissipated into the corona is then given by
\begin{equation}
f = \frac{\lc}{1+\lc}.
\end{equation}
If $\lc \gg 1$, then $f \sim 1$ since most of the gravitational energy is
dissipated into the corona and $F_{\rm d} \approx F_{\rm th}$.  Conversely,
if $\lc\ll 1$, then $f \approx 0$ since most of the gravitational energy is
dissipated directly in the disk and $F_{\rm d} \approx P_{\rm G}$. We study
models for $0.01 \le \lc \le 1000$ corresponding to $0.01\lta f \lta
1$. Although setting $l_{\rm G}(1-f)$ to unity is an arbitrary choice, the
main results presented in this paper are independent of the value. For a
given total optical depth, the self-consistent coronal temperature depends
on the ratio of the coronal compactness parameter to the intrinsic disk
compactness parameter, which is a function of $f$ (for a more detailed
discussion, see HM93).

The temperature of the disk, $\Tbb$, is held fixed, as in HM91.  The
spectral shape of the thermal radiation is given by Planck's law,
normalized such that the total flux of thermal radiation is consistent with
the self-consistent compactness parameter of the disk (including
reprocessing).  With a fixed $\Tbb$, different values of $l_{\rm d}$
correspond to different values of the coronal scale-height $z_{\rm 0}$.

With \lc\ , \taue\ , and \Tbb\ specified, the NLMC code is used to find the
resulting temperature structure, $T(z)$, the total opacity $\tauT$
(including the contribution by pairs), the internal radiation field, and
the escaping radiation field. The spectrum of escaping radiation is stored
in ten bins in $\mu$ between $0 \le \mu \le 1$, where $\mu=\cos\theta$ and
$\theta$ is the angle between the normal of the disk and the photon's
direction.  We also compute the spectrum of radiation incident onto and
reflected from the disk, also stored in ten bins in $\mu$.  As mentioned in
\S\ref{prevmodsec}, the reprocessing of radiation in the disk is computed
by interfacing a linear MC technique with our LPMC code.  We are able to
calculate the energy-integrated disk albedo, $a$, more accurately than HM93
since we do not assume the radiation field incident onto the disk field is
isotropic; the albedo does depend on the angular distribution of the
incident radiation field.  We stress that we do not simply add the
reflected component of the radiation, modified by an escape probability, to
the escaping spectrum.  Instead, we propagate the reflected component
through the corona in the same fashion as all other components of the
radiation field.  Therefore, multiple reflections and interactions of
reflected photons with the corona are taken into account.

\subsection{Generation of the models}
We have produced a grid of roughly 100 ADC models with a slab geometry.
For all cases, the spatial distributions of the energy dissipation rate,
$H(z)$, and the seed opacity distribution, $n_{\rm p}(z)$, are uniform.
For the purposes of comparing our simulations to the spectra of BHCs [which
we do in a companion paper (\cite{dove97b}, hereafter referenced as
paper~II)], we allowed the seed opacity to vary within the range $0.05 \lta
\taup \lta 2.0$ and the compactness parameter to vary within the range $0.1
\lta \lc \lta 10^3$.  Except for the soft X-ray excess, the output spectra
are not sensitive to the value of $\Tbb$, and therefore we have fixed $k\Tbb
= 200$\,\eV\ (see discussion in \S~\ref{resultssec}).

The grid is produced by sweeping the coronal compactness parameter from its
minimum to its maximum values while keeping the seed opacity fixed. After
each sweep, the seed opacity is incremented.  Each new
simulation is started using the end-state of the previous simulation.  In
this approach, CPU time is saved since the system does not have to evolve
much prior to reaching an equilibrium state when compared to starting a
simulation from scratch.  The only caveat in producing the grid is that the
relative increments in \lc\ should not be more than about three-fold; for
larger increments, the system is unable to smoothly evolve to the new
equilibrium solution since the transitional behavior will heavily disturb
the system. We have verified that the results presented here are
independent of the order in which the grid is produced (no
hysteresis). After the system has reached an equilibrium, the spectrum of
escaping radiation is binned and the self-consistent properties of the
corona are determined.

\section{Results}\label{resultssec}
\subsection{Consistency Tests and Comparisons to Linear MC Models} 
In order to test our model, we compared the spectra of escaping radiation,
produced by our {\em linear} version of the model (where we do not include
reprocessing, pair-production, annihilation, and we use uniform density and
temperature structures), with the linear MC models of \cite*{wilms97a} and
Skibo (private communication).  Both slab and spherical geometries
were considered.  In all cases, our models are in excellent
agreement with all of the linear MC models.  Additionally, we computed the
amplification factor $A$, where $A = L_{\rm out}/L_{\rm input}$, with
$L_{\rm out}$ and $L_{\rm input}$ being the luminosity (erg s$^{-1}$) of
the escaping radiation and the luminosity of the injected seed photons,
respectively.  We have compared our amplification factors with those given
by \cite*{gorecki84} for spherical geometry.  These amplification factors
agree to within 7\%, and we attribute the differences to numerical
fluctuations. Due to computational speed constraints, \cite*{gorecki84} did
not simulate many particles.  Amplification factors were also computed
using the linear MC code of \cite*{wilms97a}. Our values agreed with these
values to within 2\%.  As an additional test, we computed the average
number of scatterings a photon undergoes prior to escaping the system. We
compared these results with the analytic expectations for both the
optically thin and optically thick cases. We also compared the average
change in the photon's energy per scattering event, as well as the average
change in the square of the energy change ($\left<\epsilon^2\right>$), and
compared them to the analytic expectations for both the relativistic and
non-relativistic cases.  In all cases, our results are perfectly
consistent.

We compared the spectrum of reprocessed (i.e.\ reflected) radiation when
injecting the cold accretion disk with a power-law, to the spectra given by
\cite*{george91} and \cite*{haardt93b}.  Finally, we compared our maximum
self-consistent coronal temperature values, as a function of opacity, to
those given by \cite*{stern95b}, \cite*{poutanen96a}, and
\cite*{haardt97a}.  Although our results agree with \cite*{stern95b} for
all optical depths, there are some discrepancies with the other models for
$\tauT \lta 0.2$.  For example, for $\tauT = 0.2$, we predict an average
coronal temperature $\Tc = 148$\,\keV\ while \cite*{haardt97a} predict $\Tc
= 160$\,\keV, a discrapency of $7.5$\%.  The discrapency increases as the
opacity decreases.  We do not understand these discrepancies, and an
investigation that will address this problem is currently underway.  Since
the discrepancy is $< 10$\% for $\tauT \ge 0.2$, however, these
differences are not important for understanding the hard spectra of BHCs,
as all ``best-fit'' slab models have a total opacity $\tauT \ge 0.3$.

\subsection{Self-Consistent Thermal Properties of the Corona}
\begin{figure*}
\centerline{\psfig{file=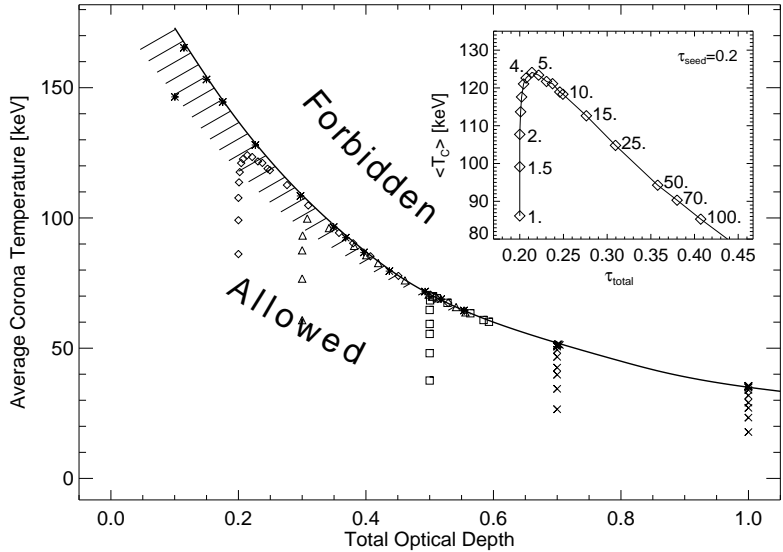,width=0.6\textwidth,angle=-0}}
\caption{Allowed temperature and opacity regime for self-consistent ADC
models with a slab geometry.  Solid line is derived from a ``fit by eye''
to the numerical results.  For a given total optical depth, temperatures
above the solid line are not possible.  The hatched region below the solid
line marks the parameter space in which the contribution to the total
opacity by pairs is significant.  For all models, the blackbody temperature
of the disk is $k\Tbb = 200$\,\eV.  Different symbols represent models with
different seed opacities. Insert: The average coronal temperature and the total
opacity for several values of \lc.}\protect\label{fig:ttau2}
\end{figure*}

In Fig.~\ref{fig:ttau2}, we show the self-consistent temperature as a
function of the total optical depth, for several values of \lc\ and seed
opacities.  
The solid line marks the maximum coronal temperature, as a
function of the total opacity, $\Tmax(\tauT)$.  This figure is similar to
Fig.~1a of HM93 and Fig.~1 of \cite*{stern95a}, but, for clarity, we
include models that are both pair-dominated and non pair dominated, showing
how the temperature evolves with increasing compactness parameter .  As
pointed out by HM93, the maximum self-consistent temperature, for a given
total opacity, is independent of the seed opacity.  However, for a given
total opacity, pair dominated models reach the maximum temperature with a
coronal compactness parameter that is much lower than the corresponding
value for non pair-dominated models.

Also in Fig.~\ref{fig:ttau2} (insert), we show how the temperature of the
corona varies with the coronal compactness parameter, starting with a seed
opacity of $\taup = 0.2$.  Note that, for small $\lc$ values, the
temperature increases as a function of \lc\ while the opacity remains nearly
constant (pair production is negligible).  Here, the Compton cooling rate
is dominated by the soft photons implicitly emitted by the cold accretion
disk ($f \ll 1$), and an increase in \lc\ corresponds to an increase in the
heating rate without much of an increase in the cooling rate.  As the
corona becomes hotter, the pair production rate increases.  Once pair
production begins to be significant, the increase in $\tauT$ results in an
increase of radiation reprocessing within the cold accretion disk.  Since
most ($\sim$ 90\%) of this reprocessed radiation is re-emitted as a thermal
blackbody, the increase in the reprocessed radiation causes a very large
increase of the Compton cooling rate within the corona.  Thus, through the
production of pairs and the increase in reprocessed radiation, the Compton
cooling rate increases more than linearly with increasing \lc. Since the
heating rate is only proportional to \lc, the coronal temperature reaches a
maximum value and then decreases with increasing \lc\ for models where the
seed electron optical depth is held constant.  We note that this $\Tc(\lc)$
relationship is due to our assumption that the intrinsic compactness
parameter of the disk is constant while \lc\ varies.  However, this
behavior does give physical insight on how the onset of pair production
forces the temperature to reach a maximum value and then decrease with
increasing values of the coronal compactness parameter.  Any model in which
the flux of the disk becomes dominated by reprocessing of coronal radiation
will exhibit this behavior.

For the models presented here, the pair-dominated models do not predict an
observable annihilation line in the escaping spectrum; thus, for a given
total opacity and coronal temperature, the pair-dominated and non
pair-dominated models predict the same spectrum of escaping radiation.  The
degeneracy of models is shown in Fig.~\ref{fig:ttau2}, where models with
different seed opacities and compactness parameters sometimes have the same
total optical depth and temperature.  For this reason, our results are
independent of the value of the intrinsic disk compactness parameter,
$l_{\rm G}(1-f)$, which we set to unity.  If this value were set to a
higher value, the Compton cooling rate would be higher, but the same
maximum temperatures would be reached with higher \lc\ values.  If $l_{\rm
G}(1-f)$ were decreased, the models in which $\lc\gg 1$ would not be
altered since the thermal emission is already dominated by reprocessed
radiation.  For models where $\lc \lta 1$, decreasing $l_{\rm G}(1-f)$,
while holding the seed opacity constant, simply reduces the Compton cooling
rate and precludes the cooler regions of Fig.~\ref{fig:ttau2} (lower left
region) from being self-consistent (i.e., reducing the value of the
intrinsic disk compactness parameter drives the solutions towards the
pair-dominated, $f\sim1$, regime).  All of these arguments have been
confirmed by numerical simulations.

The black-body temperature used for our models is $\Tbb\ = 200$\,\eV, a
value believed to be appropriate for BHCs (see paper~II).  However, the
results presented in Fig.~\ref{fig:ttau2} are insensitive to the disk
temperature.  For example, a corona with a total optical depth $\tauT =
0.25$ and a disk temperature $\Tbb = 5$\,eV has a maximum temperature
$\Tmax = 117$\,keV, compared to $\Tmax = 128$\,keV for the $\Tbb = 200$\,eV
case.  These values are consistent with the results of HM93. Therefore,
models with very low disk temperatures, which are attractive in the sense
that the thermal-excess might not be observable due to the efficient
Galactic absorption of ultraviolet radiation, have maximum coronal
temperatures that are colder than the values presented above. On the other
hand, models with a higher black-body temperature predict even larger
thermal-excesses than those corresponding to a $200$\,eV disk.

We conclude that accretion disk coronae cannot have a temperature higher
than $\sim 120$\,\keV\ if the total optical depth is $\gta 0.2$, regardless
of the corona compactness parameter, intrinsic accretion disk compactness
parameter, and the black-body temperature.  Alternatively, ADCs with a
temperature $\gta 150$\,\keV\ cannot have an optical depth greater than
$\sim 0.15$ (see Fig.~\ref{fig:ttau2}).  The range of allowed
self-consistent values of the coronal temperatures and opacities for the
slab geometry corresponds to a minimum value of the angle-averaged photon
index $\Gamma \sim 1.8$ (as determined by numerically fitting the spectra
between $5$\,\keV\ and $30$\,\keV\ with a power-law), which places severe
limitations on the applicability of these ADC models to BHCs.  However, due
to the anisotropy break (HM93), the power-law portion of the spectrum
hardens as the inclination angle decreases (see Fig.~\ref{fig:plotmufig}).
For models with $\tauT = 0.3$ and $k\Tc = 110$\,\keV\, where the
angle-averaged index $\Gamma \approx 1.8$, $\Gamma$ varies between $2.1$
and $1.7$ as the orientation varies from `edge-on' to `face-on,'
respectively.

In the accompanying paper~II, we demonstrate that ADC models with a slab
geometry cannot explain the broad-band X-ray spectrum of Cyg X-1 and other
BHCs.  The observed hard power-laws in BHCs can be explained only with
models having photon-starved sources, where Compton-cooling is less
efficient and the coronae are allowed to have higher temperatures. We find
that ADC with a spherical geometry and an exterior cold accretion disk
allow for such conditions.

\subsection{Non-Uniform Temperature Structure}
\begin{figure*}
\centerline{\psfig{file=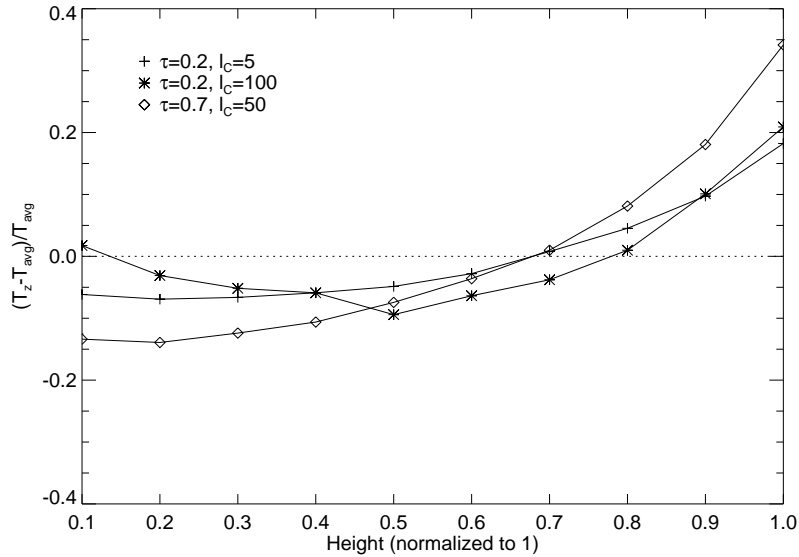,width=0.6\textwidth,angle=-0}}
\caption{Vertical temperature structure of the accretion disk corona
expressed as the relative deviation of the local temperature, $T(z)$ from
the volume averaged corona temperature $\avg{\Tc}$. Several models, with
different combinations of the seed optical depth and coronal compactness
parameter, are shown.}\protect\label{fig:cortempz}
\end{figure*}

As discussed in \S~\ref{modelsec}, our models allow for the determination
of the temperature and opacity within local cells, which allows for the
possibility of non-uniform temperature structures.  In
Fig.~\ref{fig:cortempz}, we give the relative temperature deviation
($\Delta T/\avg{\Tc}$) for assorted values of $\taup$, \Tc\, and $l_{\rm
c}$.
Note that, for optically thin models ($\tauT \lta 0.3$), the temperature is
essentially uniform, while the temperature varies up to a factor of three
for optically thick models.  For $\tauT \gta 0.5$, the region nearest the
accretion disk is always the coldest, while the region farthest away from
the disk (largest values of $z$) is the hottest.  This is easily understood
since the energy density of the radiation field decreases with increasing
height, resulting in a decrease of the Compton cooling rate with $z$.
For optically thin coronae, the energy density of the radiation field does not
vary too much with height; however, the average photon energy of the radiation
field increases with height since the spectrum of the internal
radiation field hardens with increasing height.  Therefore, since the
Compton cooling rate is proportional to both the average energy, $<\!
\epsilon \!>$, and the energy density, the minimum temperature can occur at
intermediate heights rather than always at $z=0$.

\subsection{The Spectrum of Escaping Radiation}
\subsubsection{Comparison to Uniform Models}
\begin{figure*}
\centerline{\psfig{file=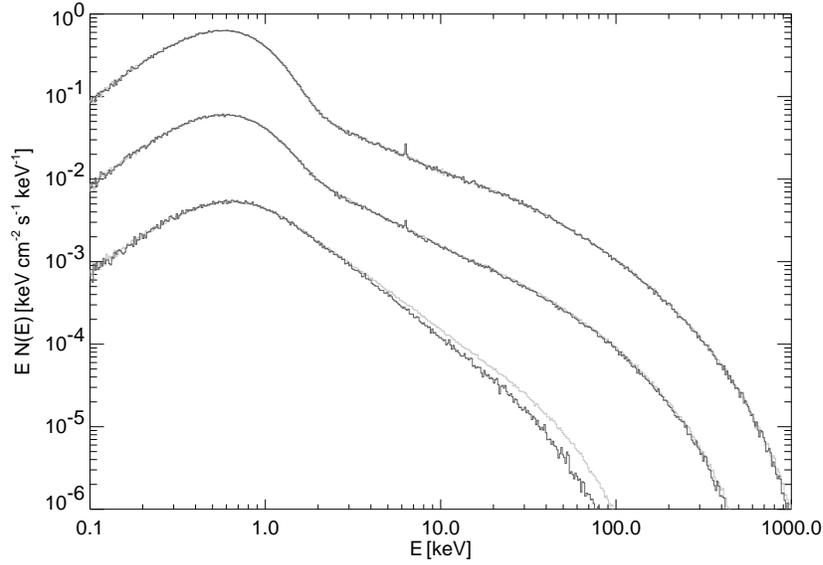,width=0.6\textwidth,angle=-0}}
\caption{A comparison of the predicted spectra of escaping radiation
from our non-linear ADC models to that of our linear models, where the
vertical temperature distribution is uniform.  From top to bottom, the
average temperature and total optical depth \{$k\avg{\Tc}(\keV),\tauT$\}
are: \{118.4, 0.25\}, \{69.8, 0.50\}, and \{15.6, 2.0\}. Black lines
represent the non-linear models, and gray lines represent the linear models
(indistinguishable for the two optically thin models). For clarity, each
spectrum is normalized to an arbitrary value.}\protect\label{fig:nlvslin}
\end{figure*}

In Fig~\ref{fig:nlvslin}, for $\tauT =$\,0.25, 0.50, and 2.0, we compare
the escaping spectra between models with a uniform temperature structure to
non-uniform models (arrising from a uniform heating distribution). 
For the optically thin models, the thermal gradients do not give rise to
any significant changes in the spectral shape of the radiation field.  In
fact, as shown in Fig~\ref{fig:nlvslin}, the spectra are virtually
indistinguishable, as the fractional difference between the uniform and
non-uniform models is no more than 1\% for the opticaly thin
models. Therefore, for optically thin ADC models, a non-uniform
distribution of the heating rate is required to sustain temperature
gradients large enough such that the spectrum of escaping radiation
significantly differs from the uniform models.  For parameters appropriate
for low mass X-ray binaries, where $\tauT \gta 1$, the spectrum of the
escaping radiation is significantly different from the corresponding
spectrum resulting from a uniform model.  The non-uniform model produces a
slightly harder high energy tail, and is caused by the small contribution
of the uppermost region of the corona, where the temperature is hotter than
the average coronal temperature.

\subsubsection{Anisotropy of the Escaping Radiation Field}\label{fmusec}
\begin{figure*}
\centerline{\psfig{file=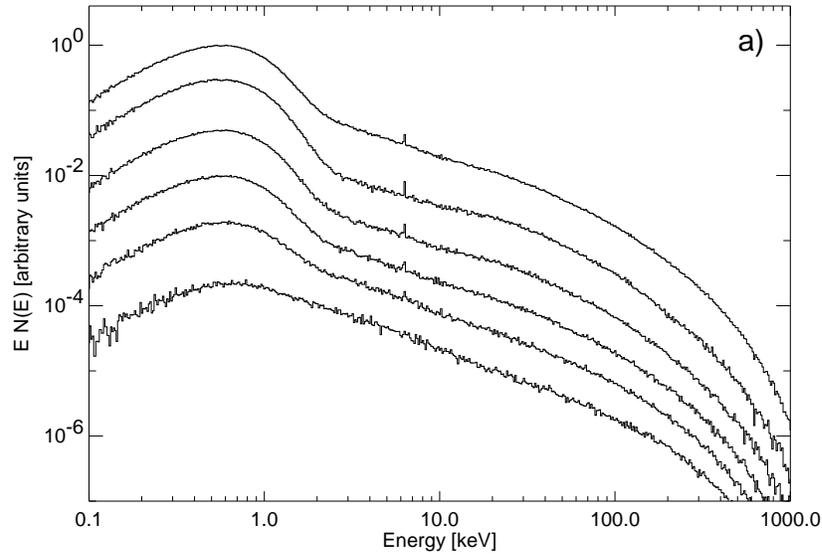,width=0.6\textwidth,angle=-0}}
\caption{The spectra of escaping radiation for several values of $\mu$,
where $\mu = \cos(\theta)$.  The uppermost plot is the spectrum averaged over
all angles.  Below, in order, $\mu = 0.9, 0.7, 0.5, 0.3$, and $0.1$.
For all cases, the average coronal temperature is $k\avg{\Tc} =
118$\,\keV, the optical depth is $\tauT = 0.25$, and the black-body
temperature of the disk is $k\Tbb = 200$\,\eV.}\protect\label{fig:plotmufig}
\end{figure*}

In Fig.~\ref{fig:plotmufig}, we show the spectrum of the escaping radiation
for several values of the inclination angle.  
We define $\mu = \cos(\theta)$, where $\theta$ is the angle between the
line of sight and the normal of the accretion disk.  As expected, the
spectral shape does not vary with respect to the inclination angle for
optically thick models.  However, for optically thin models, the spectral
shape of the escaping radiation field (and the internal radiation field)
{\em does} vary with $\mu$.  Since the effective optical depth for
radiation leaving the accretion disk is $\tau/\mu$, the fraction of
black-body radiation that escapes the system without subsequent
interactions with the coronane

decreases with decreasing $\mu$.  Consequently, the thermal-excess gets
weaker with decreasing values of $\mu$ (These arguments also apply to the
strength of the Fe~K$\alpha$ fluorescence line).  Finally, the spectrum
becomes softer as $\mu$ decreases since the magnitude of the reflection
`hump' decreases with the effective optical depth.  These
results are in agreement with HM93.  Therefore, the spectra
of ADCs that are seen `face-on' ($\mu=1$) are the hardest, but they also
have the largest amount of thermal excess.  On the other hand, ADCs viewed
``edge on'' do not have an observationally recognizable soft-excess in
their spectra, nor do they contain any reprocessing features.  If the
matter responsible for producing the ``reflection features'' is the
optically thick, cold accretion disk that is also producing the seed
photons, it does {\em not} appear possible that the spectra of ADCs can
have strong reflection features without also containing a strong soft
excess.  While a very cold accretion disk (i.e., $k\Tbb \lta 10$\,eV) would
produce thermal emission that could be ``hidden'' due to the efficient
Galactic absorption of UV radiation, it is questionable whether an
accretion disk of a BHC, in the radial region where most of the
seed-photons are produced, could be so cold (For a more detailed
discussion, see paper~II).

\subsection{Strength of Fe K$\alpha$ Fluorescence Line}
The strength of the Fe~K$\alpha$ fluorescence line for BHCs provides
stringent constraints on the properties of ADC models.  For a slab
geometry, roughly half of the Comptonized radiation field within the corona
is reprocessed within the cold accretion disk.  For parameters appropriate
for BHCs, where the radiation field is hard (having a photon index of
$\Gamma \sim 1.5$), the large amount of reprocessing gives rise to a very
large equivalent width (EW) of the Fe K$\alpha$ line.  For optically thin
models having a temperature $130 \ge k\Tc \ge 100$\,\keV\, the
angle-averaged EW of the iron line is $120\,\eV \ge {\rm EW} \ge 90\,\eV$
(for each temperature, there is a scatter of about $10$\,\eV\ due to the
range of optical depths, the uncertainty in the EW measurements, and the
statistical nature of the Monte Carlo simulations).  The predicted EWs
would be even higher than these values if a harder spectrum (i.e. a
spectral shape that describes the Cyg X-1 observations better than the
self-consistent spectrum) were incident onto the cold accretion disk.  For
example, according to simulations with our linear Monte Carlo code, a
radiation field with a power-law index of 1.5 and an exponential cutoff at
150\,\keV\ irradiated onto the cold disk will result in an EW $\sim
150$\,\eV\ for a `face-on' orientation.  Although there is a great deal of
uncertainty, the observed EW of the Fe~K$\alpha$ line of BHCs is very
small, usually having only an upper limit rather than a true detection.
For Cyg X-1, the EW is $\lta 70$\,\eV\ (\cite{ebisawa96b,gierlinski97b}).
Therefore, unless the abundance of iron is much less than solar, the slab
geometry models predict EW values that are significantly larger than the
values observed for Cyg~X-1.  As we discuss in paper~II, this problem is
not encountered with models having a sphere$+$disk geometry. For that
geometry, the accretion disk, as seen by the corona, has a covering factor
roughly least one third of that for the slab geometry, giving rise to less
reprocessing of coronal radiation (\cite{done92}).

\subsection{Comparisons to Previous Models}
\subsubsection{Hua \& Titarchuk (1995)}
\begin{figure*}
\centerline{\psfig{file=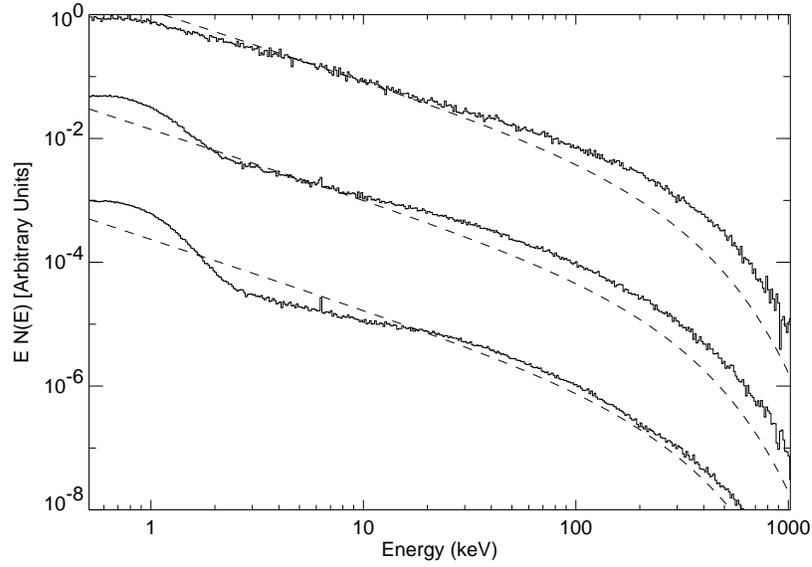,width=0.6\textwidth,angle=-0}}
\caption{Comparison of the predicted spectrum of escaping radiation with
the analytic model of Titarchuk (\protect\cite*{hua95}), for several
inclination angles.  Solid lines correspond to our model, and the hashed
lines correspond to Titarchuk's model.  From top to bottom, $\mu =$ 0.1,
0.5, and 0.9.  Again, $k\avg{\Tc} = 118$\,\keV, $\tauT = 0.25$, and $k\Tbb =
200$\,\eV.}\protect\label{fig:comptitfig}
\end{figure*}

The major difference between our models and the analytic models of
\cite*{sunyaev80} and \cite*{hua95} is that our models take into account
the reprocessing of radiation within the cold accretion disk, which gives
rise to reflection features in the spectra.  In addition, our models give a
more accurate description of the transition between the thermal black-body
(seed photons) and the Comptonized portion of the spectrum since
Titarchuk's analytical model is only valid for energies much higher than
the energy of the seed photons (nor does this model include any anisotropic
effects).  Therefore, it is expected that they will predict different
spectra of escaping radiation.  However, due to the popularity of this
model, we give a comparison (Fig.~\ref{fig:comptitfig}) of our models to
the XSPEC version of Titarchuk's slab-geometry Comptonization model
(\cite{hua95}). 
Here, the spectral shape of the seed photons for
Titarchuk's model is described by a Wien law rather than a Planckian
distribution.  Although the two models differ significantly, they
do agree for energies much higher than $k\Tbb$ if the reflection features
are ignored.  However, our high-energy tails do not decrease with
increasing energy as rapidly as the analytic model.  This disagreement has
been discussed by several other authors (\cite{stern95a} and
\cite{skibo95}). We also note that, with the current implementation of 
XSPEC (version 9.0), it is not possible to add a reflection component
to Titarchuk's Comptonization model self-consistently.

\subsubsection{Power-Law with Exponential Cutoff}
\begin{figure*}
\centerline{\psfig{file=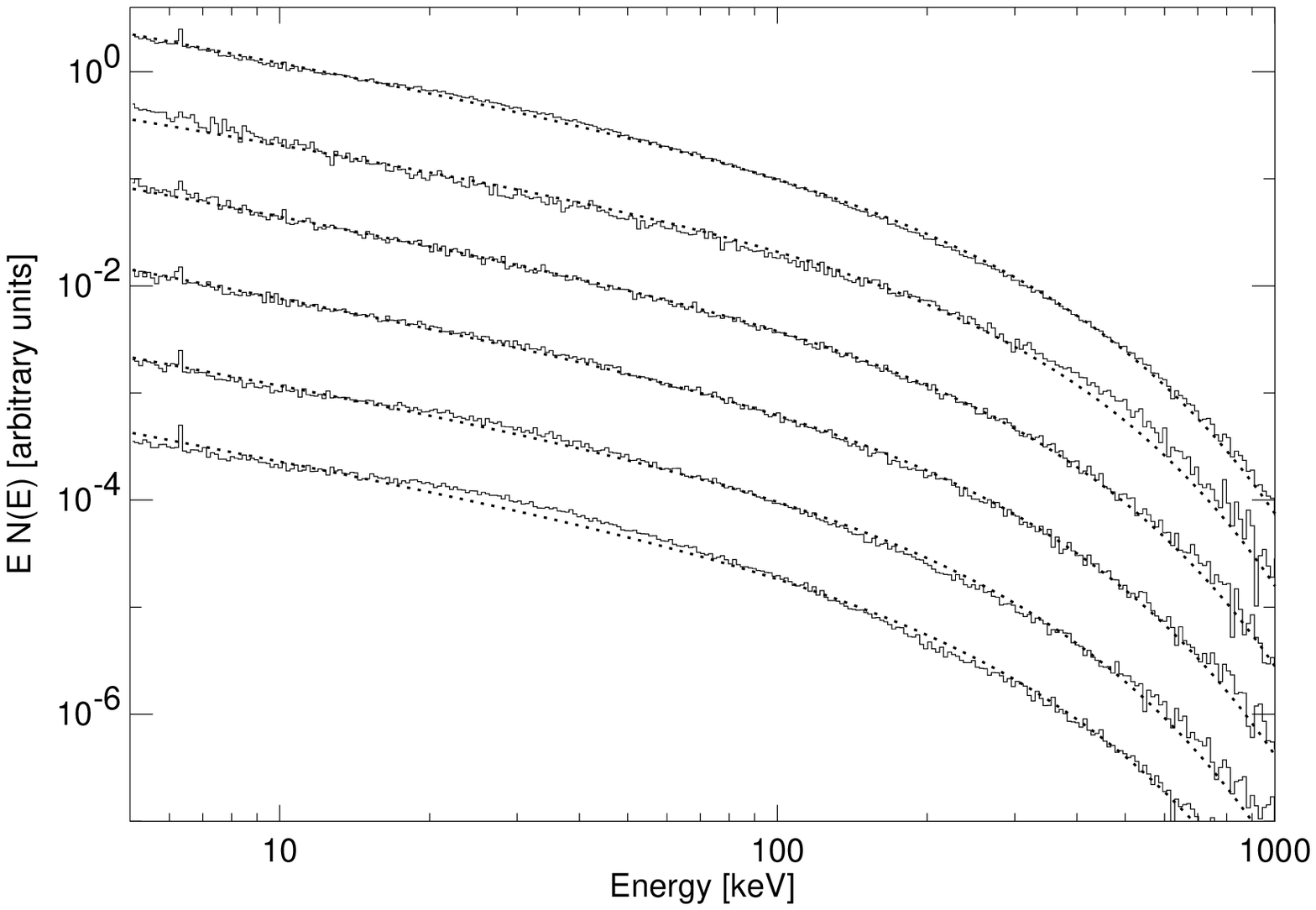,width=0.48\textwidth}%
\hfill%
\psfig{file=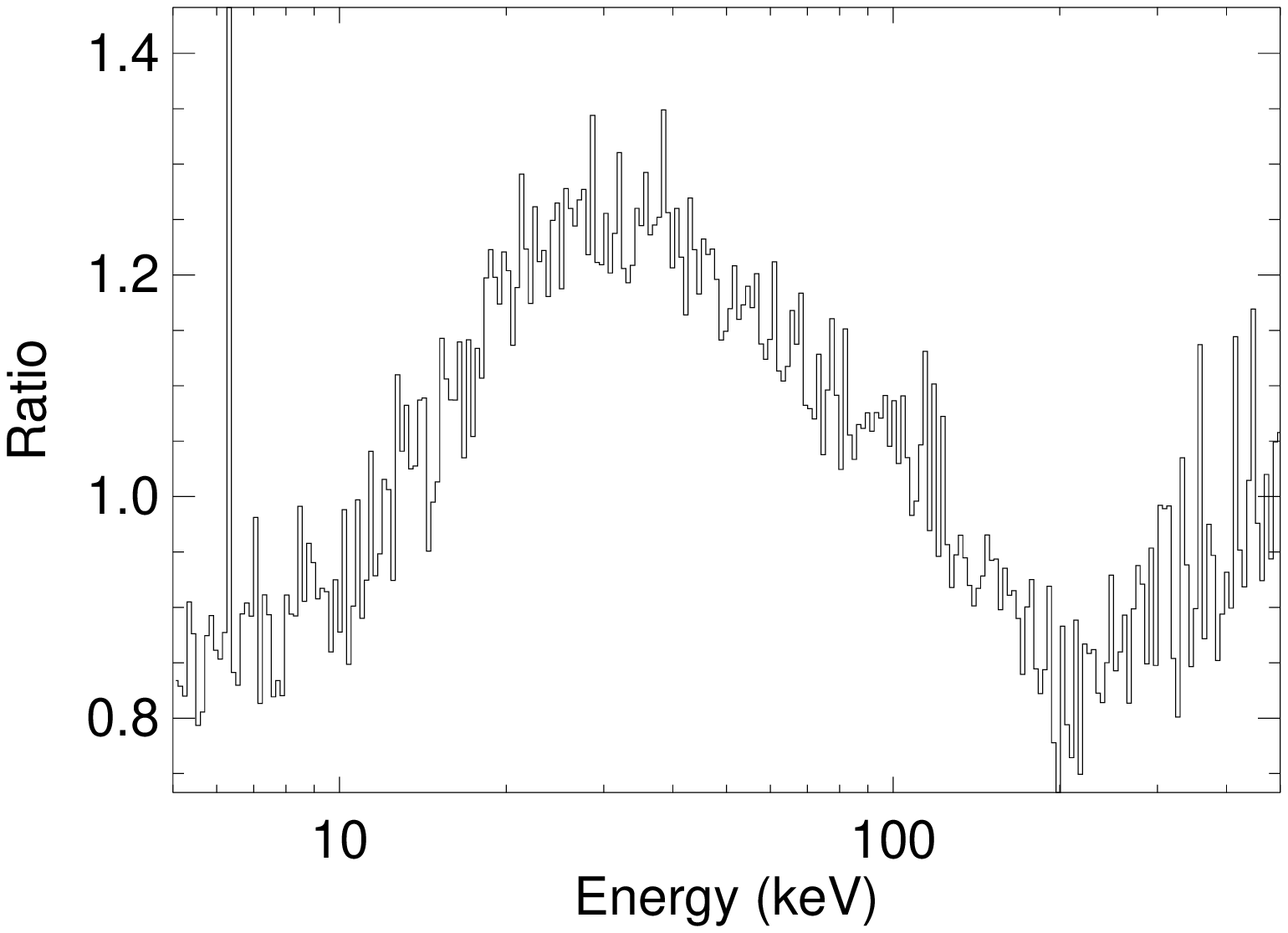,width=0.48\textwidth}}
\caption{Comparison of the predicted spectrum of escaping radiation with
spectra of the form $F(E) \propto E^{-\Gamma}\exp(-E/E_{\rm cut})$.  (a)
The uppermost spectrum is the angle-averaged spectrum; then, from top to
bottom, $\mu = $\,0.1, 0.3, 0.5, 0.7, and 0.9.  The model parameters are
the same as in Fig.~\ref{fig:comptitfig}. (b) The residuals of the
numerical model divided by the analytical model for $\mu =
0.9$.}\protect\label{fig:plexpfig}
\end{figure*}

As shown in Fig.~\ref{fig:plexpfig}a, we have attempted to describe our
simulated spectra of escaping radiation by a power-law with an exponential
cut-off,
\begin{equation}\label{eq:plexp}
N_{\rm E} = F_{\rm 0}E^{-\Gamma} \exp(-E/E_{\rm 0}),
\end{equation}
where $N_{\rm E}$ is the photon flux as a function of energy, $E$.
Although we were able to achieve good `by-eye' fits for the angle-averaged
spectrum (integrated over all values of $\mu$), we were unable to find good
fits for individual bins of $\mu$.  This problem is due to the presence of
reflection features.  In Fig.~\ref{fig:plexpfig}b we show the residuals of
our numerical model divided by the analytical model
(equation~\ref{eq:plexp}) for $\mu=0.1$.  The fits become better as the
inclination angle increases since the strength of reprocessing features
decreases with increasing angles (\S~\ref{fmusec}).  It appears to be a
coincidence that, for the angle-averaged spectrum, the reflection features
add up in such a way that a power-law is better preserved.

\subsubsection{Reflected Exponentially Cutoff Power-law}
\begin{figure*}
\centerline{\psfig{file=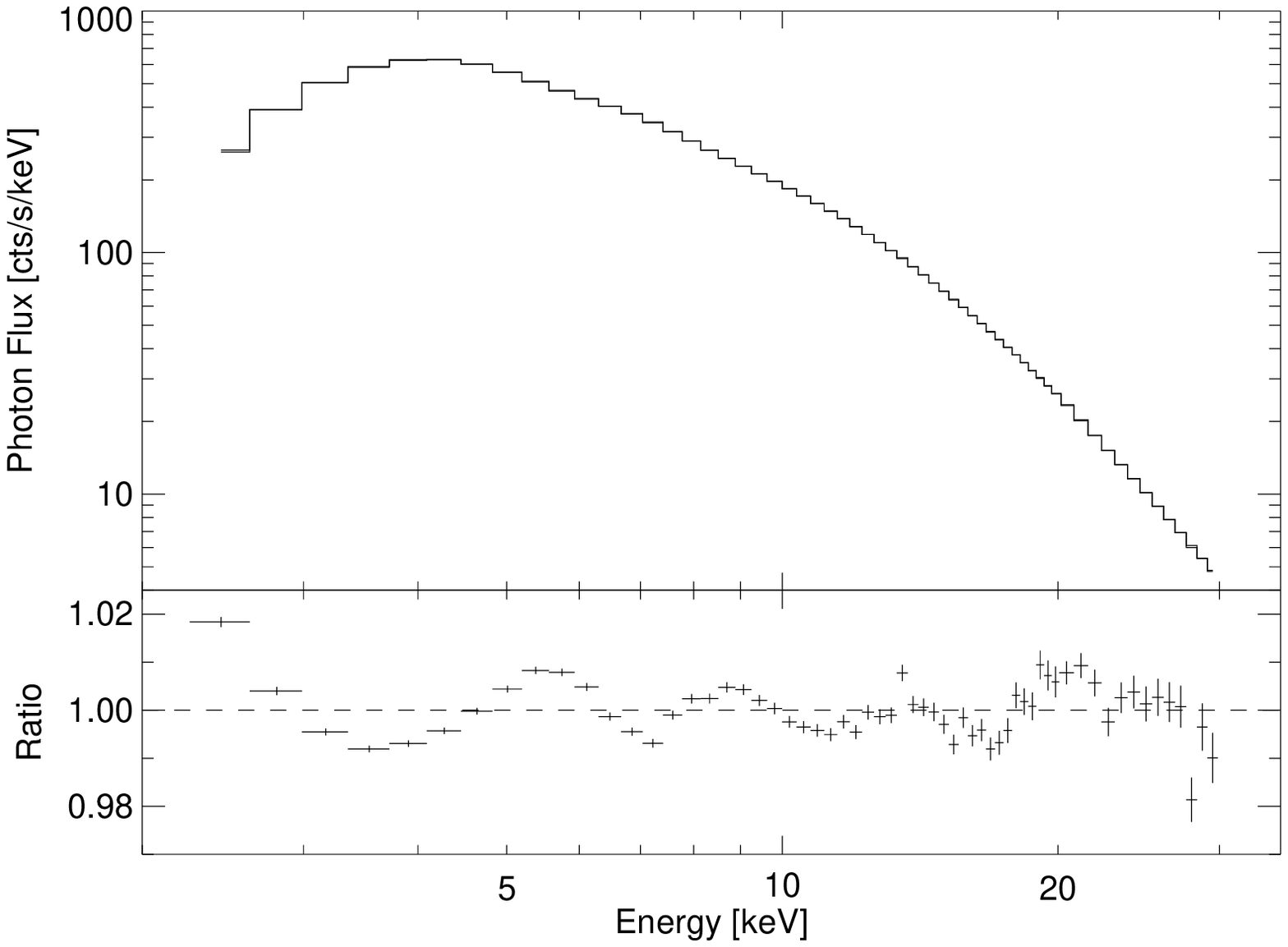,width=0.6\textwidth,angle=-0}}
\caption{Comparison of the reflected exponentially cutoff power-law model,
{\em pexrav} (\protect\cite{magdziarz95a}), to the NLMC model.  Using {\em
XSPEC} (\protect\cite{arnaud96a}), ``fake'' data was produced by folding
the NLMC model through the RXTE PCA response matrix.  The parameters of the
NLMC model are the same as in Fig.~\protect\ref{fig:comptitfig}. A Gaussian
at $6.4$\,\keV\ and a blackbody distribution with $k\Tbb = 200$\,\eV\ have
been added to the {\em pexrav} model.  In the {\em pexrav} model, an
isotropic radiation field given by $N(E) \propto E^{-\Gamma}\exp[-E/E_c]$
is reflected by a slab of cold matter.  The ``best-fit'' reflection model
resulted in an equivalent width of the Fe~K$\alpha$ line $EW \approx
100$\,\eV, $\Gamma \approx 2.0$, $E_{\rm c} = 400$\,\keV, and the covering
fraction of the disk $f=0.48$.  Galactic absorption has been taken into
account by both models using {\em wabs} (using the Morrison and McCammon
cross-sections) and assuming $N_{\rm H} = 6\times 10^{21}$
cm$^{-2}$.}\protect\label{fig:fitkotelp}
\end{figure*}

Using XSPEC, we have compared our model to the reflected exponentially
cutoff power-law model {\em pexrav} (\cite{magdziarz95a}). Using the RXTE
PCA response matrix, we simulated a $10^4$ second observation using {\em
fakeit} and our NLMC model, where $k\Tbb = 200$\,\eV, $k\avg{\Tc} =
118$\,\keV, $\tauT = 0.25$, and $\mu = 0.9$. As shown in
Fig.~\ref{fig:fitkotelp}, 
We fit these `data' with a superposition of a
$200$\,\eV\ blackbody, a Gaussian line at $6.4$\,\keV, and {\em pexrav}.
Our best fit yielded a photon index of $2.0$, a cutoff energy $E_{\rm c}
= 400$\,\keV, and the relative reflection parameter $f=0.48$. 
We froze the abundance parameter to unity, and set $\mu = 0.9$.  It is
clear that there is much more structure in our NLMC model than in the
reflection model, showing a pitfall of assuming that a Comptonized,
un-reflected spectrum can be described by an exponentially cutoff
power-law. It is also interesting that the best fit covering fraction is
$f<0.5$ given that the physical covering fraction for the slab geometry is
$f=1$.

\section{Conclusions}\label{concsec}
We present accretion disk corona models where the radiation field, the
temperature, and the total opacity of the corona are determined
self-consistently.  We take into account the coupling between the corona
and the accretion disk by including reprocessing of radiation in the
accretion disk.

The range of self-consistent temperatures and total opacities
places severe limitations on the model's applicability in explaining the
X-ray spectra of BHCs.  The maximum corona temperature for models having
$\tauT>0.3$ is $k\Tc \approx 110$\,\keV.  The corresponding angel-averaged,
$5$\,\keV\ - $30$\,\keV\ photon index is $\Gamma \gta 1.8$, a value too large
to explain the observed hard spectra of BHCs (with typical photon power-law
indices of $\sim 1.5$).  Even models with a `face-on' orientation predict
photon indeces of $\Gamma \gta 1.7$.  The spectra (for energies $E \gta
5$\,\keV) of BHCs, most notably Cyg~X-1, have been adaquately described by
linear ADC models where $\tau \sim 0.3-0.5$ and $kT \sim 150$\,\keV\ (e.g.\
\cite{haardt93a,titarchuk94b}).  All of these previous ADC models are not
self-consistent, for they used temperature and opacity values that are
outside the allowed region.

In addition, as described in paper~II, attempts to describe the broad-band
spectrum of Cyg~X-1 have proven to be much more difficult than modeling the
spectrum over a smaller energy range.  This difficulty is due to the fact
that optically thin ADC models always predict a very large thermal excess,
a feature not found in the observations, unless the inclination angle is
nearly `edge-on.' For the case of Cyg X-1, these high inclination angles
probably can be ruled out since it appears that the inclination angle is
between 32$^{\circ}$ and 40$^{\circ}$ (Ninkov, Walker, \& Yang, 1987).
Additionally, models with edge-on orientations predict that no reprocessing
or reflection features will be present in the spectra.  Optically thick
models, which also predict a small thermal excess, have a self-consistent
coronal temperature that is much too cold to explain the spectra of BHCs.
Due to galactic absorption, the predicted thermal excess decreases with
decreasing values of the black-body temperature. However, a colder
accretion disk results in a higher Compton cooling rate, yielding even
lower coronal temperatures than the values corresponding to a 200\,\eV\
disk.

\section*{ACKNOWLEDGEMENTS}
We thank M.~Nowak and M.~Maisack for the many useful discussions and
B.~Stern for the development of the first version of the NLMC code. This
work has been financed by NSF grants AST91-20599, AST95-29175, INT95-13899,
NASA Grant NAG5-2026 (GRO Guest Investigator program), DARA grant 50 OR
92054, and by a travel grant to J.W.  from the DAAD

\end{multicols}
\end{document}